\documentclass[sigconf]{acmart}
\usepackage{graphicx}
\usepackage{booktabs}
\usepackage{amsmath}
\usepackage{multirow}
\usepackage{float}
\usepackage{natbib}
\usepackage{array}
\usepackage{placeins}
\usepackage{enumitem}
\usepackage{algorithm}
\usepackage{algpseudocode}

\AtBeginDocument{%
  }

\copyrightyear{2026}
\acmYear{2026}
\setcopyright{cc}
\setcctype{by}
\acmConference[KDD '26]{Proceedings of the 32nd ACM SIGKDD Conference on Knowledge Discovery and Data Mining V.1}{August 09--13, 2026}{Jeju Island, Republic of Korea}
\acmBooktitle{Proceedings of the 32nd ACM SIGKDD Conference on Knowledge Discovery and Data Mining V.1 (KDD '26), August 09--13, 2026, Jeju Island, Republic of Korea}
\acmPrice{}
\acmDOI{10.1145/3770854.3785686}  
\acmISBN{979-8-4007-2258-5/2026/08}  

\begin{document}

\title{From Memorization to Creation: Evaluating the Cognitive Depth of LLM‑Generated Educational Questions}

\author{Xiaolong Wang}
\authornote{Xiaolong Wang and Zhe Zhao contributed equally to this work. This work was done during Xiaolong Wang's internship at Squirrel Ai Learning.}
\affiliation{%
  \institution{Squirrel Ai Learning}
  \city{Shanghai}
  \country{China}
}
\email{wmumu8533@gmail.com}

\author{Zhe Zhao}
\authornotemark[1]  
\affiliation{%
  \institution{City University of Hong Kong}
  \city{Hong Kong}
  \country{China}
}
\email{zzhao26-c@my.cityu.edu.hk}

\author{Song Lai}
\affiliation{%
  \institution{City University of Hong Kong}
  \city{Hong Kong}
  \country{China}
}
\email{songlai2\_c@my.cityu.edu.hk}

\author{Chaoli Zhang}
\authornotemark[2]  
\affiliation{%
  \institution{Zhejiang Normal University}
  \institution{Squirrel Ai Learning}
  \city{Jinhua}
  \country{China}
}
\email{chaolizcl@zjnu.edu.cn}

\author{Zijie Geng}
\affiliation{%
  \institution{University of Science and Technology of China}
  \city{Hefei}
  \country{China}
}
\email{ustcgzj@mail.ustc.edu.cn}

\author{Yu Tong}
\affiliation{%
  \institution{Wuhan University}
  \city{Wuhan}
  \country{China}
}
\email{yutchina02@gmail.com}

\author{Ye Wei}
\authornote{Corresponding authors.}
\affiliation{%
  \institution{City University of Hong Kong}
  \city{Hong Kong}
  \country{China}
}
\email{ye.wei@cityu.edu.hk}

\author{Qingsong Wen}
\affiliation{%
  \institution{Squirrel Ai Learning}
  \city{Bellevue}
  \country{USA}
}
\email{qingsongedu@gmail.com}

\renewcommand{\shortauthors}{Xiaolong Wang et al.}

\begin{abstract}
While LLMs show promise in automating educational content creation, their ability to generate questions that stimulate higher-order thinking remains understudied. This work evaluates six widely-used LLMs through a Bloom's Taxonomy lens, focusing on their capacity to transcend rote memorization and achieve cognitive leaps. Using a hybrid human--AI evaluation protocol, we generate and analyze 20{,}700 questions across computer science, K--12 math, and social-science domains. Key contributions include: (1) a fine-grained prompting strategy that reduces question repetitiveness by 24.45\% for Qwen2.5-7B-Instruct, and increases the proportion of higher-order cognitive level outputs by 11.53\% for InternLM3-8B-Instruct; (2) quantitative metrics for cognitive shift intensity (CogShift) and category drift, revealing InternLM3's superior performance in multi-level transitions; (3) an interpretability analysis revealing metric-level correlations that enhance the transparency of Chain-of-Thought prompting. Our findings highlight the importance of cognitive-aware prompt design and provide benchmarks for deploying LLMs in personalized learning systems.
\end{abstract}



\maketitle

\section{Introduction}

The rapid advancement of artificial intelligence, particularly the emergence of large language models (LLMs), is transforming the future of online education~\citep{scaria2024automated, chen2024dr}. LLM-based automated question generation has become a cornerstone of intelligent educational systems due to its efficiency, scalability, and potential for personalization~\citep{elkins2024how, li2024planning}. This technology enables the rapid generation of a vast number of questions based on key knowledge points in the curriculum, significantly alleviating the repetitive workload of teachers in question design while providing accurate, dynamic support for adaptive learning platforms~\citep{wei2022chain, chen2024empowering}.

\begin{figure}[!t]
    \centering
    \includegraphics[width=0.98\linewidth]{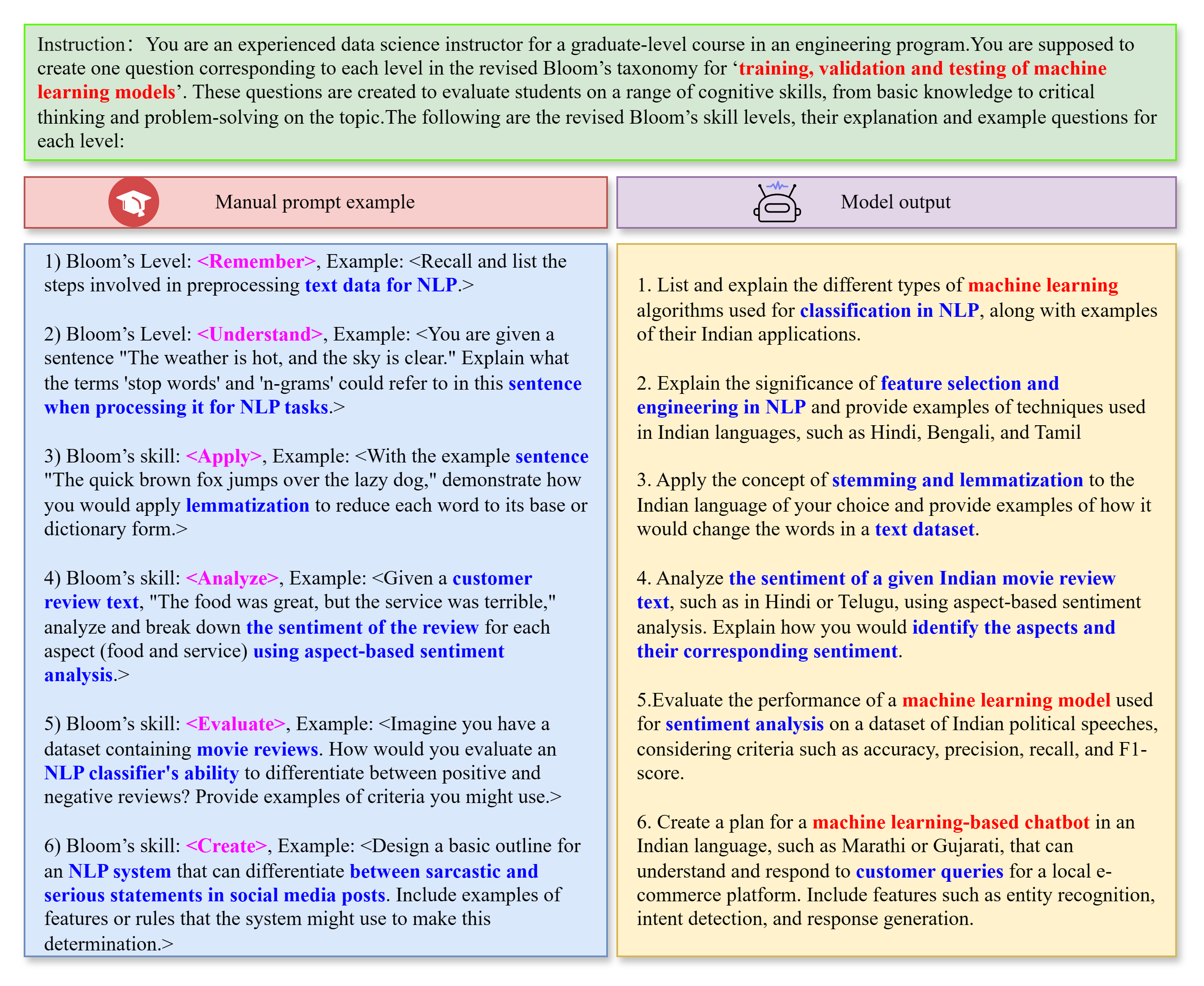}
    \caption{
    \textbf{An illustration of current limitations in LLM-generated question assessment.} Red highlights indicate \textit{knowledge units}, and blue highlights mark \textit{expert-crafted examples}. The generated output shows notable lexical overlap with exemplars but poor alignment with the intended knowledge unit. The examples are sourced from the publicly available dataset introduced by Scaria et al.~\cite{scaria2024automated}.}
    \label{fig:gen_bias_issue}
\end{figure}

Existing research primarily focuses on the quality and knowledge relevance of generated questions, often overlooking a fundamental goal in education: the rational progression of cognitive levels~\citep{kojima2022large, liu2025from}. The true value of education lies not only in the transmission of knowledge but also in guiding learners through cognitive leaps—from ``memorizing facts'' to ``creating solutions''~\citep{scaria2024automated, li2024planning}. This ability to make cognitive transitions reflects the depth of human learning and serves as a critical metric for assessing the efficacy of educational tools~\citep{gong2022khanq, mucciaccia2025automatic}. For instance, a regression from higher to lower cognitive dimensions can result in questions that fall below a student's actual cognitive level, thus lacking sufficient challenge~\citep{tobler2022impact, daheim2024stepwise}. Conversely, cognitive shifts between adjacent dimensions promote incremental learning and stimulate cognitive development~\citep{fedyk2023how, li2024planning}. Additionally, a jump from lower to higher cognitive levels encourages deep thinking but also induces a higher cognitive load~\citep{muse2023pre, park2024large}.

Thus, accurately capturing cognitive level transitions and incorporating them into the evaluation of question generation is becoming an essential area of research for large models in educational content creation~\citep{muse2023pre, chen2024dr, arvan2023linguistic}. By refining methods to assess cognitive shifts, we can enhance the adaptability of LLMs in personalized education, ensuring that generated questions appropriately challenge learners at various stages of their cognitive development~\citep{scaria2024automated, yuan2022selecting}.

Research in educational psychology indicates that effective learning is based on a hierarchical progression of cognitive abilities, Bloom’s taxonomy serving as a systematic framework for this process~\citep{anderson2001taxonomy, scaria2024automated}. The framework categorizes cognitive abilities into six levels: remembering, understanding, applying, analyzing, evaluating, and creating, emphasizing the need for incremental breakthroughs to enhance students' overall cognitive abilities~\citep{anderson2001taxonomy, li2024planning, chen2024empowering}. This hierarchical approach is critical for structuring educational experiences that foster cognitive growth and facilitate deep learning~\citep{elkins2024how, kojima2022large}. Furthermore, it highlights the importance of designing learning experiences that encourage transitions between adjacent cognitive levels, promoting an environment conducive to progressive skill development~\citep{wei2022chain, yuan2022selecting}.

However, as shown in Figure \ref{fig:gen_bias_issue}, current question generation and evaluation methods based on LLMs have notable limitations: (1) Knowledge units do not always align with example questions, leading to insufficient knowledge relevance in generated questions; (2) The evaluation process focuses primarily on the consistency of cognitive categories, neglecting potential cognitive shifts across different cognitive levels. This limitation prevents effective adaptation to learners’ actual cognitive levels and fails to foster deep thinking or facilitate cognitive leaps.

To address these challenges, we present a Bloom’s-Taxonomy-driven framework for question generation and evaluation that pivots from a knowledge-point perspective to a cognitive-leap paradigm. Our contributions can be summarized as follows:
\begin{itemize}
    \item \textbf{Precision alignment and generation control:} High-quality exemplar questions authored by experienced educators are precisely matched with their corresponding knowledge units, the target Bloom level for each question is explicitly specified, and contextual learning is leveraged to mitigate cognitive regression during generation.
    
    \item \textbf{Comprehensive quantitative assessment:} We devise an integrated metric suite—question usefulness, knowledge coverage, cognitive consistency, and cognitive shift—that objectively evaluates both the quality and the cognitive depth of generated questions.
    
    \item \textbf{Empirical interpretability insights:} Experiments demonstrate strong correlations between knowledge identification and coverage, as well as between question-classification scores and category consistency, thereby enhancing the interpretability of Chain-of-Thought prompting.
\end{itemize}

\begin{table*}[!t]
\centering
\small
\caption{Distribution of questions across datasets by Bloom's levels and associated knowledge units.}
\label{tab:data_distribution}
\renewcommand{\arraystretch}{1.1}
\setlength{\tabcolsep}{4.5pt}
\begin{tabular}{l c c c c c c c c}
\toprule
\multirow{2}{*}{\textbf{Dataset}} &
\multirow{2}{*}{\textbf{Total}} &
\multicolumn{6}{c}{\textbf{Bloom's Levels}} &
\multirow{2}{*}{\textbf{Knowledge Units}} \\
\cmidrule(lr){3-8}
& & \textbf{Remember} & \textbf{Understand} & \textbf{Apply} & \textbf{Analyze} & \textbf{Evaluate} & \textbf{Create} & \\
\midrule
Computer Science & 1404 & 231 & 103 & 362 & 288 & 98 & 322 & 12 \\
K--12 Math       &  438 & --  & --  & 438 & --  & --  & --  & 19 \\
Social Science   &  100 & 17  & 18  & 17  & 16  & 15  & 17  & 5 \\
\bottomrule
\end{tabular}
\end{table*}

\section{Related Work}

\subsection{Educational Question Generation}

Recent progress in LLMs has revealed strong zero-shot and few-shot reasoning capabilities, particularly under structured prompting strategies. Kojima et al.~\cite{kojima2022large} demonstrated that LLMs can exhibit competent reasoning performance in zero-shot settings when guided by carefully designed prompts. Wei et al.~\cite{wei2022chain} introduced chain-of-thought prompting to elicit step-by-step reasoning, significantly improving model accuracy on complex tasks. Shi et al.~\cite{shi2023prompt} further optimized the prompt space to enhance few-shot reasoning through automated search and selection. These capabilities have spurred applications in educational content generation.

Kokku et al.~\cite{kokku2018augmenting} integrated AI systems into classroom environments to support personalized instruction and adaptive feedback. Laban et al.~\cite{laban2022quiz} proposed a system to assist educators in quiz design by leveraging automated question generation technologies. Wang et al.~\cite{wang2024book2dial} generated teacher-student interactions from textbooks, enabling scalable development of educational chatbots. To ensure cognitive relevance, many studies incorporate Bloom’s taxonomy into the question generation pipeline.

Elkins et al.~\cite{elkins2024how} aligned prompting strategies with Bloom’s cognitive levels to produce more pedagogically valid quiz items. Bartel et al.~\cite{bartel2023applying} discussed the practical challenges of implementing cognitive learning principles, such as Bloom’s taxonomy, in real-world educational research. D'Silva and Matlen~\cite{dsilva2023embedding} emphasized equity and rigor in the design of cognitive learning interventions, advocating for inclusive AI-powered instruction. To improve the controllability and fidelity of generated content, various techniques have been proposed.

Qiu and Chen~\cite{qiu2025knowledge} presented a knowledge graph reasoning model for adaptive testing, supporting precise concept-level control. Feng and He~\cite{feng2025rgr} enhanced logical form generation for QA by integrating structured knowledge into LLMs. Ravikiran et al.~\cite{ravikiran2025teemil} developed a framework to estimate MCQ difficulty in Indic languages, facilitating language-sensitive question calibration. Chen et al.~\cite{chen2022grow} proposed a grow-and-clip strategy to extract concise yet informative evidence, improving answer explanation quality in educational settings. Liao et al.~\cite{liao2025skintern, liao2025awakening} improved controllability and interpretability in LLM-based question generation by embedding symbolic knowledge and activating internal representations.

\begin{table}[t]
\footnotesize
\setlength{\tabcolsep}{4.5pt}
\centering
\caption{Comparison of our framework with representative instruction/data synthesis and educational QA systems. DS = data synthesis, Bloom = Bloom alignment, Cog = cognitive evaluation, Cov = knowledge coverage, Human = human evaluation, Edu = educational focus. ``\checkmark'' indicates that the aspect is explicitly addressed; ``--'' means it is not a primary focus.}
\label{tab:rw_comparison}
\resizebox{\linewidth}{!}{%
\begin{tabular}{lcccccc}
\toprule
Method & DS & Bloom & Cog & Cov & Human & Edu \\
\midrule
Self-Instruct~\cite{wang2023selfinstruct}      & \checkmark & -- & -- & -- & \checkmark & -- \\
MAGPie~\cite{xu2025magpie}                     & \checkmark & -- & -- & -- & --        & -- \\
Auto-Q~\cite{constantinides2025autoq}          & \checkmark & -- & -- & -- & --        & -- \\
TeacherQuiz~\cite{elkins2024how}               & --         & \checkmark & -- & -- & \checkmark & \checkmark \\
AutoEduQ~\cite{scaria2024automated}            & --         & \checkmark & -- & \checkmark & \checkmark & \checkmark \\
Dr.\ Academy~\cite{chen2024dr}                 & --         & \checkmark & -- & \checkmark & \checkmark & \checkmark \\
PlanQG~\cite{li2024planning}                   & --         & -- & -- & -- & --        & \checkmark \\
AutoMCQ~\cite{mucciaccia2025automatic}         & --         & -- & -- & -- & \checkmark & \checkmark \\
QuizDesign~\cite{laban2022quiz}                & --         & -- & -- & -- & \checkmark & \checkmark \\
\textbf{Our work}                              & --         & \checkmark & \checkmark & \checkmark & \checkmark & \checkmark \\
\bottomrule
\end{tabular}%
}
\end{table}

\subsection{Evaluation Benchmarks}

Evaluation of educational question generation typically relies on expert annotation or Bloom-style classification. Scaria et al.~\cite{scaria2024automated} conducted a comprehensive benchmark study across multiple LLMs, assessing their ability to generate questions aligned with Bloom’s cognitive levels. Elkins et al.~\cite{elkins2024how} evaluated LLM-generated quiz items using Bloom’s taxonomy to examine the consistency between intended and realized cognitive skills. Anderson and Krathwohl~\cite{anderson2001taxonomy} provided the revised Bloom’s taxonomy, which continues to serve as the foundation for cognitive-level assessment in education. Chen et al.~\cite{chen2024dr} introduced Dr.\ Academy, a benchmark treating LLMs as teaching agents and evaluating generated questions across consistency, relevance, coverage, and representativeness. Gong et al.~\cite{gong2022khanq} released the KhanQ dataset, offering cognitively annotated questions that support deep-question generation and fine-grained evaluation. 

Park et al.~\cite{park2024large} developed a zero-shot framework to estimate question difficulty using LLMs as student simulators, modeling their performance across Bloom levels. Muse et al.~\cite{muse2023pre} demonstrated that domain-specific pretraining on scientific texts enhances the depth and quality of generated educational content. Yuan et al.~\cite{yuan2022selecting} proposed sampling and ranking techniques to select high-quality questions from LLM outputs, improving educational alignment. Tobler et al.~\cite{tobler2022impact} investigated the role of prior knowledge in narrative-based learning, highlighting the importance of matching question complexity with learner readiness. Arvan et al.~\cite{arvan2023linguistic} analyzed cognitive load in instructional dialogues, revealing how linguistic features influence comprehension. Fedyk and Ray~\cite{fedyk2023how} advocated integrating machine learning interpretability with cognitive psychology to inform experimental hypothesis generation.

Despite these efforts, most existing benchmarks lack mechanisms to jointly assess knowledge relevance and cognitive depth, and seldom capture shifts in Bloom-level complexity. Table~\ref{tab:rw_comparison} situates our framework among representative instruction-tuning and educational QA systems along six methodological dimensions. To address this gap, we propose a unified evaluation framework that quantifies question quality, Bloom-level consistency, and cognitive drift in LLM-generated educational content.

\begin{table*}[t]
  \caption{Evaluation rubric for assessing generated questions based on expert exemplars.}
  \label{tab:evaluation}
  \centering
  \small
  \renewcommand{\arraystretch}{1.1}
  \setlength{\tabcolsep}{8pt}  
  \begin{tabular}{
    >{\centering\arraybackslash}p{2.5cm}  
    >{\centering\arraybackslash}p{2.5cm}  
    >{\raggedright\arraybackslash}p{8.5cm}  
  }
    \toprule
    \textbf{Evaluation Criteria} & \textbf{Rubric Item} & \textbf{Description} \\
    \midrule
    \multirow{3}{*}{Question Usability}
      & Unique     & Is the generated question highly similar to the original question? (yes/no) \\
      & Readable   & Is the question clearly stated? (yes/no) \\
      & Answerable & Can students reasonably answer the question? (yes/no) \\
    \midrule
    Category Consistency & Bloom’s Level   & Remember, Understand, Apply, Analyze, Evaluate, Create \\
    \midrule
    Knowledge Coverage   & Knowledge Unit & [Knowledge units list] \\
    \bottomrule
  \end{tabular}
\end{table*}

\section{Methodology}

We propose a reliable and fully automated evaluation framework for systematically assessing the performance of LLMs in educational question generation. The framework applies two prompting strategies to multiple LLMs to generate questions across diverse knowledge units and Bloom’s cognitive levels, and employs automated evaluation to analyze their quality and cognitive alignment, providing insight into model strengths and limitations at different cognitive stages.

\subsection{Dataset}

We employ two structured educational datasets in computer science and K--12 mathematics to ensure domain consistency in question content and knowledge units, supporting expert-level evaluation. The full list of knowledge topics is provided in Table~\ref{tab:dataset_topics}. These datasets form the foundation for evaluating LLM performance across cognitive and content dimensions.

For computer science, we use the Data Structure subset comprising 1{,}406 questions across 12 core knowledge units, such as stacks, queues, binary search trees, hashing, and recursion, etc. Cognitive levels are automatically annotated using a Bloom-aligned verb list, with each question classified into one of the six Bloom levels: remember, understand, apply, analyze, evaluate, and create~\citep{zaman2024dataset}. The distribution of these levels is shown in Table~\ref{tab:data_distribution}.

To assess LLMs' ability to generate high-quality questions across different knowledge areas under a fixed cognitive level, we incorporate a K--12 mathematics dataset from the Squirrel AI intelligent tutoring system\footnote{\url{https://squirrelai.com/}}. This platform is widely deployed across primary and secondary education in China and features a large-scale, adaptive item bank. We sample 438 mathematics application problems that align specifically with the ``apply'' level in Bloom's taxonomy, covering 19 knowledge units, such as geometry, functional reasoning, and combinatorics, etc. All problems are explicitly mapped to the ``apply'' cognitive level to ensure alignment with the evaluation framework.

To probe generalization beyond STEM domains and to support small-scale human evaluation of our automated metrics, we further curate a social-science subset of 100 exam-style questions from a publicly available repository of educational exam questions\footnote{\url{https://figshare.com/articles/dataset/Exam_Question_Datasets/22597957}}. Each item is manually assigned to one of the six Bloom levels and mapped to its corresponding social-science knowledge unit, and the overall distribution is incorporated into Table~\ref{tab:data_distribution}.

\subsection{Prompting Strategies}

As illustrated in Figure~\ref{fig:prompt}, we propose two prompting strategies, Chain-of-Thought (CoT) and Fine-Grained prompting (FGP), to enhance knowledge alignment and cognitive control in the generation of LLM-based educational questions. In both, LLMs act as Bloom’s taxonomy experts within the computer science or K-12 mathematics domains. We guide generation through exemplar-based few-shot prompting and a consistent output format.

CoT prompting guides the model to first identify relevant knowledge units and cognitive levels, and then generate questions accordingly. This step-by-step reasoning improves logical consistency and cognitive alignment while enhancing interpretability.

Fine-grained prompting directly specifies the target knowledge unit and Bloom level in the input, eliminating intermediate reasoning. This approach ensures stricter generation control, particularly in the K–12 Math dataset from Squirrel AI, where the Bloom level is fixed to `apply' to focus evaluation on knowledge alignment.

\subsection{Evaluation Methodology}

We establish a fully automated, multi-dimensional evaluation framework covering question usability, category consistency, cognitive deviation, and knowledge coverage. We evaluate the quality of generated questions from three complementary dimensions, as summarized in Table~\ref{tab:evaluation}:

\begin{itemize}
    \item \textbf{Usability}: Evaluates whether a generated question is pedagogically valid, based on three binary sub-criteria:
    \begin{itemize}
        \item \textit{Unique}: Checks for substantial semantic variation from the exemplar. Superficial edits (e.g., rewording or reordering) are deemed repetitive. \textit{E.g., ``What is a binary search tree?'' vs. ``Explain the concept of a binary search tree and provide an example''---the latter differs in cognitive depth and is considered valid.}
        
        \item \textit{Readable}: Assesses grammaticality and clarity. Ambiguous, convoluted, or poorly structured questions are marked as unreadable.
        
        \item \textit{Answerable}: Determines whether a question can be reasonably answered based on provided information. \textit{E.g., asking for time and space complexity of unspecified algorithms is unanswerable.}
    \end{itemize}
    
    \item \textbf{Category Consistency}: Measures alignment between the generated question's cognitive level and the annotated Bloom’s category. A mismatch indicates a failure to meet the intended cognitive target.
    
    \item \textbf{Knowledge Coverage}: Evaluates whether the question reflects all key concepts within the annotated knowledge unit(s). Coverage is granted only when core curricular content is accurately represented.
\end{itemize}

Additionally, we introduce two auxiliary tasks to support interpretability analysis: (1) \textbf{question classification}, where the model predicts the Bloom level of the original question; and (2) \textbf{knowledge identification}, where the model identifies the knowledge unit associated with the original question.

A question is considered \textit{useful} only if it satisfies all three criteria: uniqueness, readability, and answerability. We automatically assess these dimensions using OpenAI’s GPT-3.5-Turbo-0613, GPT-4o-mini, and GPT-4.1-mini. To ensure evaluation reliability, we apply majority voting across models and discard low-agreement cases (Krippendorff’s alpha < 0.7).

After filtering out non-useful questions, we perform higher-order evaluations focusing on Bloom’s category consistency and knowledge unit coverage. Let the set of valid (i.e., useful) questions be denoted as \( Q' = \{ q'_1, \dots, q'_N \} \), where each generated question \( q'_i \) corresponds to an original question \( q_i \), annotated with its original Bloom category \( C_{\mathrm{ori},i} \), while the generated question is automatically labeled as \( C_{\mathrm{new},i} \). 

To measure category consistency between the original and generated questions, we define an indicator function:

\begin{equation}
\delta(C_{\mathrm{ori},i}, C_{\mathrm{new},i}) =
\begin{cases}
1, & C_{\mathrm{ori},i} = C_{\mathrm{new},i}; \\
0, & \text{otherwise}.
\end{cases}
\label{eq:delta}
\end{equation}

Based on this, we compute the overall category consistency (CatCons) as:

\begin{equation}
\mathrm{CatCons} = \frac{1}{N} \sum_{i=1}^{N} \delta(C_{\mathrm{ori},i}, C_{\mathrm{new},i}).
\label{eq:catcons}
\end{equation}

When \( \delta(C_{\mathrm{ori},i}, C_{\mathrm{new},i}) = 0 \), it indicates a categorical mismatch, warranting further analysis of cognitive-level transitions. 

Bloom’s cognitive process dimension is inherently structured as a hierarchical taxonomy, where each level represents an increasing degree of cognitive complexity and abstraction~\cite{anderson2001taxonomy}. To quantify such level transitions in a principled way, we assign ordinal weights to the six Bloom categories—\textit{Remember}, \textit{Understand}, \textit{Apply}, \textit{Analyze}, \textit{Evaluate}, and \textit{Create}—such that \( W_{\mathrm{ori},i},\ W_{\mathrm{new},i} \in \{1, 2, 3, 4, 5, 6\} \). This encoding reflects the widely accepted sequential structure of Bloom’s taxonomy and provides a numerically tractable basis for assessing shifts in cognitive levels.

Accordingly, the cognitive shift for a generated question is computed as:

\begin{equation}
\Delta_i = W_{\mathrm{new},i} - W_{\mathrm{ori},i}.
\label{eq:deltai}
\end{equation}

We categorize the shift direction and magnitude as follows:

\begin{itemize}
    \item $\Delta_i \geq 2$: \textbf{Cognitive Leap}, indicating a significant increase in cognitive depth.
    \item $\Delta_i \leq -2$: \textbf{Cognitive Regression}, indicating a substantial decrease in complexity.
    \item $|\Delta_i| = 1$: \textbf{Cognitive Drift}, representing a small deviation to adjacent levels.
\end{itemize}

To measure the overall extent of misalignment across the dataset, we define the \textit{Cognitive Shift Score} (CogShift) as:

\begin{equation}
\mathrm{CogShift}
= \frac{1}{N} \sum_{i=1}^{N} \left(1 - \delta(C_{\mathrm{ori},i}, C_{\mathrm{new},i})\right) \cdot |\Delta_i|,
\label{eq:cogshift}
\end{equation}

where $N$ is the number of evaluated questions after usability filtering and $\delta(C_{\mathrm{ori},i}, C_{\mathrm{new},i}) = 1$ if the predicted Bloom level matches the original target and $0$ otherwise. This metric not only captures whether a cognitive mismatch occurs but also quantifies the severity and direction of such shifts, thereby offering a fine-grained assessment of the model's cognitive control.

We also compute the proportion of each cognitive shift type (leap, regression, drift) and report their respective total shift scores to reflect the magnitude of change in each category.

To evaluate knowledge coverage, let \( K_{\mathrm{ori}} \) and \( K_{\mathrm{new}} \) denote the sets of knowledge units in the original and generated questions, respectively. We define knowledge coverage (KnowCov) as:

\begin{equation}
\mathrm{KnowCov} = \frac{|K_{\mathrm{ori}} \cap K_{\mathrm{new}}|}{|K_{\mathrm{ori}}|}.
\label{eq:knowcov}
\end{equation}

This reflects the extent to which generated questions preserve the original knowledge units. Combined, CatCons, CogShift, and KnowCov provide a comprehensive assessment of cognitive alignment and knowledge retention, supporting quantitative model diagnosis in educational contexts.

\section{Experiments}

\begin{table*}[!t]
\centering
\small
\caption{Performance of different LLMs on question uniqueness, readability, and answerability across three settings: Computer Science with CoT, Computer Science with FGP, and K–12 Math with FGP.}
\label{tab:results}
\renewcommand{\arraystretch}{1.1}
\setlength{\tabcolsep}{6pt}
\begin{tabular}{lccc|ccc|ccc}
\toprule
\multirow{2}{*}{Model} & \multicolumn{3}{c}{Comp. Sci. \& COT} & \multicolumn{3}{c}{Comp. Sci. \& FGP} & \multicolumn{3}{c}{K12 Math \& FGP} \\
\cmidrule(lr){2-4} \cmidrule(lr){5-7} \cmidrule(lr){8-10}
& Unique & Readable & Answerable & Unique & Readable & Answerable & Unique & Readable & Answerable \\
\midrule
GLM-4-9B         & 90.77 & 100    & 99.86 & 99.50 & 100    & 99.93 & 97.71 & 100 & 100 \\
Qwen2.5-7B       & 74.56 & 99.93  & 99.93 & 99.01 & 99.86  & 99.93 & 96.79 & 100 & 100 \\
Baichuan2-7B     & 85.31 & 99.15  & 96.68 & 88.89 & 99.74  & 99.47 & 97.03 & 100 & 98.63 \\
Meta-Llama-3-8B  & 97.20 & 99.14  & 98.20 & 98.02 & 99.36  & 98.87 & 96.57 & 100 & 97.94 \\
InternLM3-8B     & 94.81 & 99.86  & 99.79 & 98.94 & 100    & 99.93 & 97.07 & 100 & 99.76 \\
Spark Max        & 96.96 & 100    & 99.93 & 94.70 & 100    & 99.79 & 97.48 & 100 & 100 \\
\bottomrule
\end{tabular}
\end{table*}

\begin{table*}[!t]
\centering
\caption{LLM performance on question usefulness (Usability), Bloom-level consistency (CatCons), and knowledge coverage (KnowCov) under CoT and FGP prompting in computer science and K–12 math.}
\label{tab:llm_performance}
\renewcommand{\arraystretch}{1.1}
\setlength{\tabcolsep}{6pt}
\begin{tabular}{lccc|ccc|cc}
\toprule
\multirow{2}{*}{Model} & \multicolumn{3}{c}{Comp. Sci. \& COT} & \multicolumn{3}{c}{Comp. Sci. \& FGP} & \multicolumn{2}{c}{K12 Math \& FGP} \\
\cmidrule(lr){2-4} \cmidrule(lr){5-7} \cmidrule(lr){8-9}
& Usability & CatCons & KnowCov & Usability & CatCons & KnowCov & Usability & KnowCov\\
\midrule
GLM-4-9B          & 90.40 & 37.17 & 85.45 & 99.43 & 51.42 & 92.82 & 97.71 & 85.13 \\
Qwen2.5-7B        & 74.56 & 40.09 & 83.98 & 98.94 & 58.29 & 92.14 & 96.79 & 81.88 \\
Baichuan2-7B      & 88.45 & 36.69 & 75.97 & 82.91 & 35.86 & 90.20 & 95.66 & 81.51 \\
Meta-LLaMA3-8B    & 96.05 & 32.31 & 70.46 & 97.10 & 44.24 & 91.84 & 94.51 & 83.52 \\
InternLM3-8B      & 91.33 & 32.53 & 76.26 & 99.08 & 42.89 & 89.56 & 96.83 & 56.10 \\
Spark3.5-Max      & 96.89 & 36.57 & 78.03 & 94.48 & 47.46 & 90.42 & 97.48 & 82.15 \\
\bottomrule
\end{tabular}
\end{table*}

\label{sec:experiments}

\subsection{Settings}

To systematically evaluate the generative capabilities and adaptability of LLMs in educational contexts, we selected six widely adopted models: GLM4-9B-Chat~\citep{du2024glm4}, Qwen2.5-7B-Instruct~\citep{yang2025qwen25}, Baichuan2-7B-Chat~\citep{baichuan2023}, InternLM3-8B-Instruct~\citep{internlm2024}, Meta-LLaMA3-8B-Instruct-Chinese~\citep{dubey2024llama3}, and Spark3.5-Max\footnote{\url{https://xinghuo.xfyun.cn/}.}. Among them, all models except Spark3.5-Max are open-source and allow public access to weights and inference APIs. These models exhibit substantial diversity in terms of architecture, training corpus composition, and alignment strategy, thereby forming a representative spectrum for comparative evaluation.

To balance output diversity and generation stability, all models were configured with a temperature of 0.9 and a maximum generation length of 500 tokens. This setup aims to control sampling variability while ensuring syntactic completeness and semantic coherence in the generated questions.

We used two complementary datasets spanning computer science and K–12 mathematics to facilitate a dual-perspective evaluation along cognitive depth and knowledge breadth. The computer science dataset encompasses all six levels of Bloom’s taxonomy and provides sufficient question volume, making it suitable for assessing cognitive alignment and multi-level transfer capabilities. For this task, each model generated 1,406 questions under both the CoT and FGP strategies, resulting in 16,872 questions in total.

In contrast, the K–12 math dataset contains fewer questions but covers 19 distinct knowledge units, with all questions aligned to the ``Apply'' level in Bloom’s taxonomy. This design enables the evaluation of each model’s ability to generate diverse questions across a wide range of knowledge units while operating under a fixed cognitive constraint. Each model generated 438 questions under the FGP strategy, yielding 2,628 math questions overall. The contrast in cognitive granularity and knowledge coverage between the two tasks provides a robust foundation for analyzing LLM performance across educational dimensions. 

Finally, to examine generalization beyond STEM domains and to enable small-scale human validation of our automated metrics, each model additionally generates questions for the 100-item social-science subset under both CoT and FGP prompting configurations, yielding 1{,}200 additional questions that complement the main computer science and K--12 mathematics experiments.

\subsection{Main results}

\textbf{RQ1: How do prompting strategies affect the usefulness of educational questions generated by LLMs based on expert exemplars?} We investigate whether LLMs can generate high-quality educational questions based on expert-designed exemplars. As shown in Table~\ref{tab:results}, \textit{GLM-4-9B} consistently outperforms other models on the computer science dataset, achieving 99.50 in uniqueness, 100 in readability, and 99.93 in answerability under FGP.

Compared to CoT, FGP brings notable improvements in uniqueness across several models. \textit{Qwen2.5-7B} increases by 24.45 points (from 74.56 to 99.01), \textit{InternLM3-8B} by 4.13 points, and \textit{Baichuan2-7B} by 3.58 points. Answerability also improves slightly for \textit{Baichuan2-7B} (+2.79). These results suggest that FGP enhances both question diversity and alignment with instructional intent through finer control over generation.

However, FGP is not universally better. \textit{Spark Max}, for example, shows a 2.26-point drop in uniqueness under FGP, indicating that strict prompting may sometimes reduce generative flexibility.

On the K12 mathematics dataset (FGP only), \textit{GLM-4-9B} again performs best, achieving 99.47 in answerability and maintaining high readability and uniqueness (100 and 97.71, respectively). Notably, all models demonstrate perfect or near-perfect readability and answerability, suggesting that factual correctness and linguistic clarity are easier to satisfy in well-scoped math problems. However, uniqueness varies more significantly across models—for example, \textit{Llama3-8B} scores only 96.57 in uniqueness, indicating potential over-reliance on exemplars. 

These findings have important implications for AI-powered educational systems. Most general-purpose language models, lacking domain-specific fine-tuning or alignment through reinforcement learning, often struggle to generate content that aligns precisely with instructional objectives. In this context, prompting strategies such as FGP---which provide explicit guidance on knowledge targets and output constraints---can significantly improve the pedagogical relevance and practical utility of generated content.

\textbf{RQ2: Given that LLMs can generate useful questions, can their outputs accurately align with the target Bloom cognitive levels and cover the intended knowledge units?} As shown in Table~\ref{tab:llm_performance}, FGP significantly improves the usability of generated questions across all models. In the computer science task, the usability score of \textit{Qwen2.5-7B} increases from 74.56\% to 98.94\%, while \textit{GLM-4-9B} and \textit{InternLM3-8B} reach 99.43\% and 99.08\%, respectively. In the K-12 mathematics task, all models achieve high usability under FGP, with \textit{Spark Max} scoring the highest (97.48\%) and \textit{Baichuan2-7B} slightly lower (95.66\%). These results indicate that FGP consistently enhances the linguistic clarity and answerability of generated content across domains and model families.

After filtering out unusable questions, we further assess model performance in terms of CatCons and KnowCov—two pedagogically essential dimensions. In the computer science domain, \textit{Qwen2.5-7B} achieves an 18.20-point improvement in CatCons, and \textit{Spark Max} improves by 10.89 points. Regarding KnowCov, \textit{LLaMA-3-8B} achieves the largest gain (21.38 points), followed by \textit{GLM-4-9B} and \textit{InternLM3-8B}, which improve by 7.37 and 13.30 points, respectively.

\begin{figure*}[tbp]
    \centering
    \includegraphics[width=0.95\linewidth]{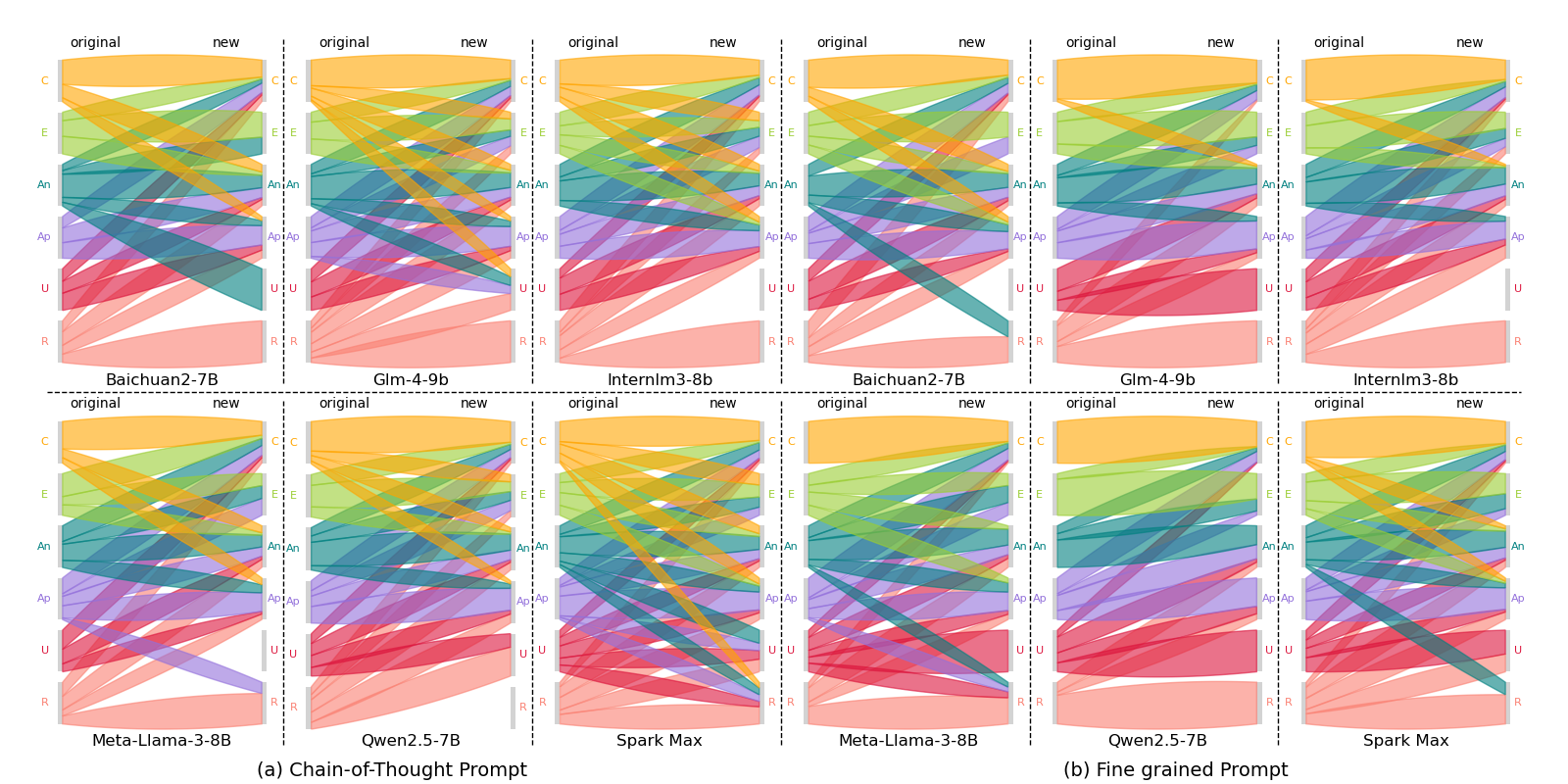}
    \caption{Cognitive level transitions of LLM-generated questions under CoT and FGP. Each subplot visualizes the Bloom-level shift between original and generated questions for a given model. Thicker bands indicate higher transition frequency.}
    \label{fig:bloom_alignment}
\end{figure*}

In the K–12 mathematics task, models exhibit substantial variation in knowledge coverage. \textit{GLM-4-9B}, \textit{Qwen2.5-7B}, and \textit{Spark Max} maintain KnowCov scores above 82\%, whereas \textit{InternLM3-8B} falls to 56.10\%, reflecting weaker knowledge alignment and a stronger tendency toward generation diversity.

Figure~\ref{fig:bloom_alignment} illustrates the cognitive transition paths under both prompting strategies. With CoT prompting, most models exhibit frequent regressions (e.g., Ap $\rightarrow$ U, An $\rightarrow$ R) and non-adjacent shifts, suggesting weak control over cognitive targeting. In contrast, FGP leads to more structured transitions, with most shifts confined to adjacent levels (e.g., U $\rightarrow$ An, An $\rightarrow$ E). \textit{InternLM3-8B} and \textit{GLM-4-9B} effectively suppress regressions, exhibiting highly stable cognitive trajectories, while \textit{Qwen2.5-7B} and \textit{LLaMA-3-8B} display clear upward transitions, reflecting stronger responsiveness to cognitive intent.

\begin{table}[t]
\centering
\footnotesize
\caption{Cognitive shift distributions (regression, drift, leap) across LLMs in computer science under Chain-of-Thought (CoT) and Fine-Grained prompting (FGP).}
\label{tab:shift_distribution}
\renewcommand{\arraystretch}{1.1}
\setlength{\tabcolsep}{2.5pt}
\begin{tabular}{lccc|ccc}
\toprule
\multirow{2}{*}{Model} & \multicolumn{3}{c}{Comp. Sci. \& CoT} & \multicolumn{3}{c}{Comp. Sci. \& FGP} \\
\cmidrule(lr){2-4} \cmidrule(lr){5-7}
& CogReg & CogDrift & CogLeap & CogReg & CogDrift & CogLeap \\
\midrule
GLM-4-9B       & 18.54 & 33.42 & 48.05 &  9.55 & 31.49 & 58.96 \\
Qwen2.5-7B     & 12.38 & 35.21 & 52.41 &  2.27 & 39.79 & 57.94 \\
Baichuan2-7B   & 22.24 & 33.12 & 44.64 & 19.54 & 27.49 & 52.96 \\
Meta LLaMA3-8B & 17.75 & 27.38 & 54.87 & 11.95 & 30.28 & 57.77 \\
InternLM3-8B   & 15.99 & 31.39 & 53.63 &  5.68 & 29.16 & 65.16 \\
Spark3.5-Max   & 28.59 & 33.26 & 38.16 & 14.41 & 32.75 & 52.84 \\
\bottomrule
\end{tabular}
\end{table}

Cognitive-level consistency and knowledge coverage reflect a model’s ability to construct coherent trajectories of reasoning and domain understanding. Structured prompting strategies constrain the cognitive transition paths and knowledge anchoring points of generated questions, enhancing the model’s controllability and intervention adaptability in instructional environments.

\textbf{RQ3: What cognitive shift patterns emerge when LLM-generated questions deviate from the intended Bloom category?} CogLeap emerges as the most prevalent deviation pattern across models and prompting strategies, indicating that LLMs tend to generate questions at higher-than-intended cognitive levels (Table~\ref{tab:shift_distribution}). Under CoT prompting, the CogLeap proportion exceeds 44\% across all models, with \textit{LLaMA-3-8B} reaching 54.87\%, \textit{InternLM3-8B} at 53.63\%, and \textit{GLM-4-9B} and \textit{Baichuan2-7B} at 48.05\% and 44.64\%, respectively. FGP further amplifies this upward bias: CogLeap increases by 11.53 percentage points for \textit{InternLM3-8B}, 10.91 points for \textit{GLM-4-9B}, 14.68 points for \textit{Spark Max}, 5.53 points for \textit{Qwen2.5-7B}, and 8.32 points for \textit{Baichuan2-7B}.

Meanwhile, FGP effectively reduces undesired downward deviations. For example, the CogReg rate drops by 8.99 percentage points for \textit{GLM-4-9B}, 10.31 points for \textit{InternLM3-8B}, and 14.18 and 10.11 points for \textit{Spark Max} and \textit{Qwen2.5-7B}, respectively. In terms of CogDrift, \textit{Baichuan2-7B} and \textit{InternLM3-8B} exhibit reductions of 5.63 and 2.23 points, respectively, indicating improved control over adjacent-level fluctuations.

\begin{figure*}[htbp]
    \centering
    \includegraphics[width=0.95\linewidth]{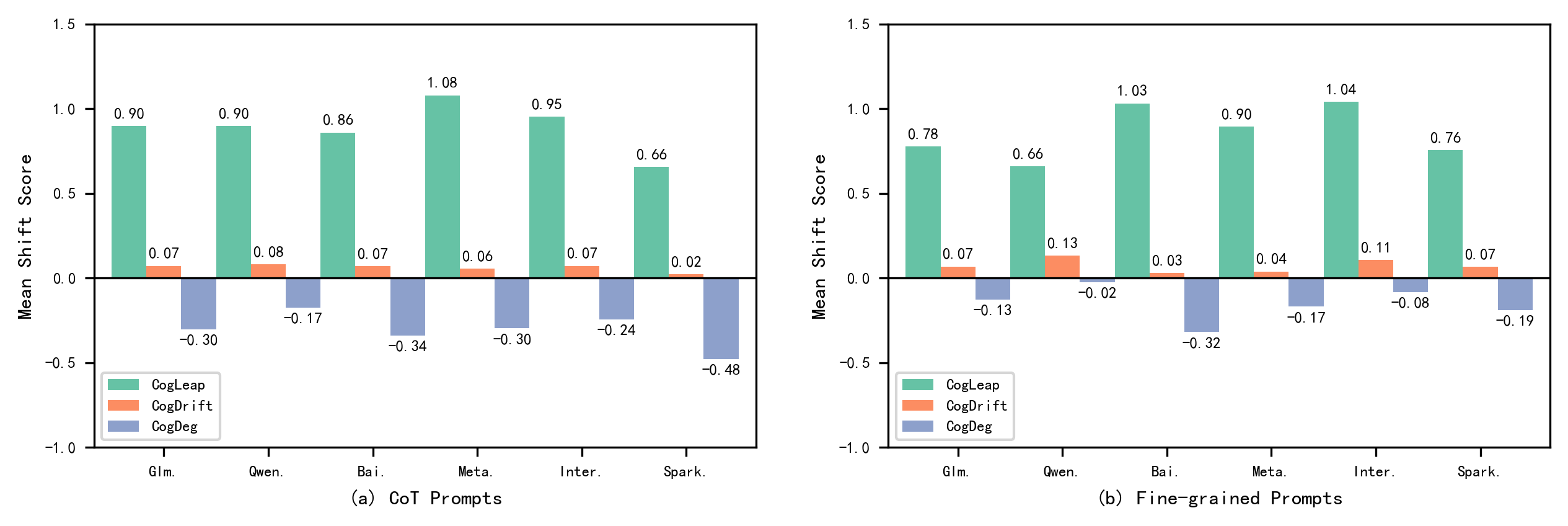}
    \caption{Mean shift score for cognitive leap, drift, and regression across LLMs under CoT and FGP.}
    \label{fig:shift_intensity}
\end{figure*}

Figure~\ref{fig:shift_intensity} further illustrates the magnitude of each shift type. \textit{LLaMA-3-8B}, \textit{InternLM3-8B}, and \textit{GLM-4-9B} accumulate the highest CogLeap scores, reflecting frequent and substantial upward transitions. In contrast, \textit{Qwen2.5-7B} maintains a moderate level of CogDrift, suggesting more stable cognitive output across tasks.

Effective instruction requires content to align with learners’ cognitive stages, gradually increasing in complexity. Models that generate questions at consistently low levels may limit learning value, while erratic shifts or abrupt leaps can disrupt cognitive flow or induce overload. Our results show that prompting strategies with cognitive control (e.g., FGP) significantly improve structural coherence and alignment, and lay a methodological foundation for progressive instruction design, targeted interventions, and adaptive learning path generation in AI-driven education.

\begin{figure*}[htbp]
  \centering
  \begin{minipage}[t]{0.36\textwidth}
    \includegraphics[width=\linewidth]{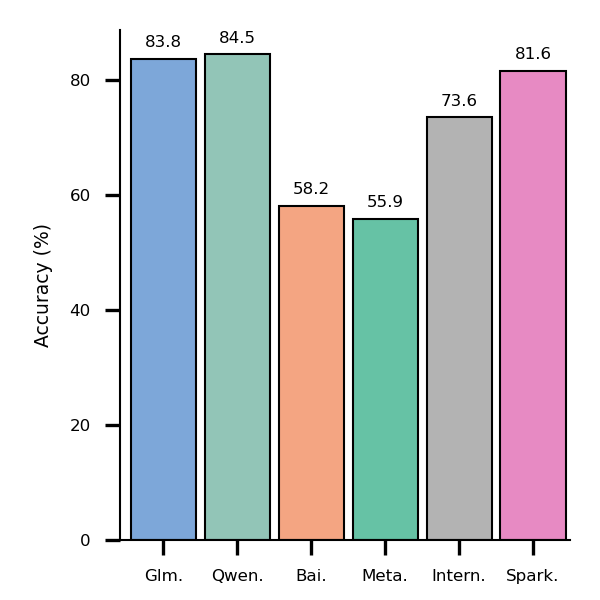}
    \caption{Model performance on the knowledge identification task, measured by accuracy in identifying the target knowledge unit to support coverage evaluation.}
    \label{fig:knowledge_identification}
  \end{minipage}
  \hspace{0.01\textwidth}  
  \begin{minipage}[t]{0.54\textwidth}
    \includegraphics[width=\linewidth]{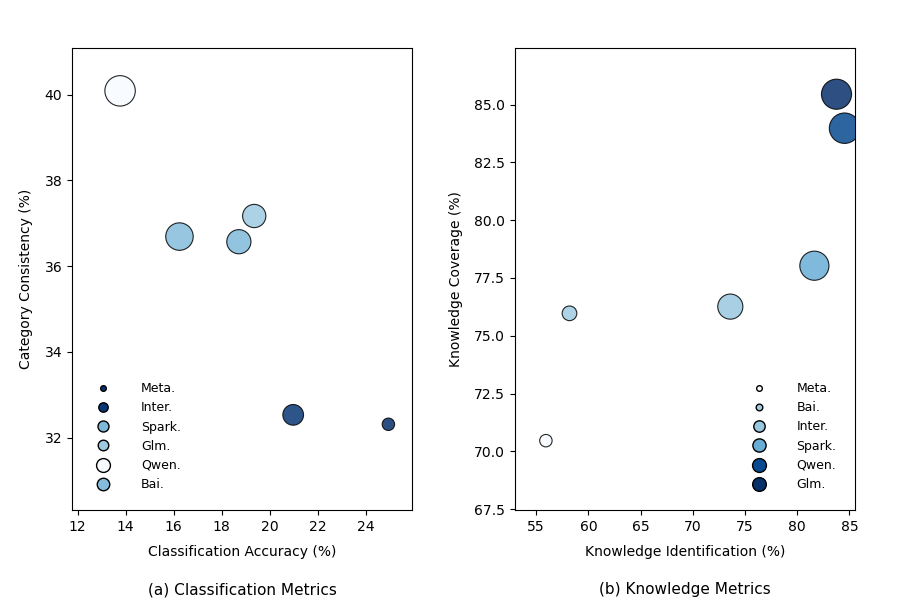}
    \caption{Correlations among evaluation metrics. (a) Bloom-level classification accuracy vs. category consistency. (b) Knowledge identification accuracy vs. knowledge coverage. Each point represents a model, with size indicating performance rank.}
    \label{fig:correlation}
  \end{minipage}
\end{figure*}

\textbf{RQ4: What is the relationship between question classification performance and category consistency under Chain-of-Thought prompting?} Counterintuitively, we observe a clear negative correlation: models with higher classification accuracy tend to exhibit lower category consistency. As shown in Figure~\ref{fig:class}, most models perform best on the ``Apply'' level, while accuracy drops markedly for higher-order categories such as ``Analyze'' and ``Create,'' which are often misclassified as lower levels like ``Understand'' or ``Apply.'' Figure~\ref{fig:correlation}(a) further quantifies this trend, confirming an inverse relationship between classification accuracy and Bloom-level consistency.

This phenomenon reflects a structural mismatch between the generative preferences and classification capacity of current LLMs in educational tasks. Regardless of whether CoT or FGP is used, LLMs consistently prefer generating semantically rich and abstract questions, reflecting an inherent tendency toward higher cognitive levels. In contrast, classification relies on shallow cues such as trigger verbs and surface syntax, which fail to capture the underlying cognitive intent and reasoning depth—leading to systematic underestimation of question complexity.

This misalignment does not result from prompt design flaws, but rather from a fundamental asymmetry between generation and understanding: while LLMs demonstrate strong abstraction and conceptual structuring in generation, their recognition mechanisms lack the depth required for reliable cognitive-level identification. Consequently, even under well-constrained prompting, LLMs struggle to model and control the cognitive objectives defined by Bloom's taxonomy, revealing a persistent gap in cognitive representation.

\textbf{RQ5: What is the relationship between knowledge identification performance and knowledge coverage in chain-of-thought prompting?} There is a strong positive correlation: higher knowledge identification accuracy is associated with improved knowledge coverage. As shown in Figure~\ref{fig:knowledge_identification}, GLM-4-9B (84.5\%), Qwen2.5-7B (83.8\%), and Spark Max (81.6\%) achieve the highest identification accuracies, with corresponding coverage rates of 85.45\%, 83.98\%, and 78.03\%. In contrast, Baichuan2-7B and Meta-LLaMA3-8B score below 60\% in identification and under 76\% in coverage, revealing consistent alignment deficiencies.

Figure~\ref{fig:correlation}(b) further confirms this monotonic relationship, with high-performing models concentrated in the upper-right region and underperforming ones in the lower-left. The alignment between identification and coverage is consistent across models, suggesting that identification quality plays a key role in knowledge-level fidelity.

The correlation indicates that knowledge identification can function as an implicit control signal for knowledge alignment. This is particularly valuable in real-world scenarios lacking explicit annotations, enabling scalable, label-free control over instructional content generation. Such an approach may enhance alignment robustness in practical educational systems.

\textbf{RQ6: How well does our Bloom-driven generation and evaluation framework generalize from STEM to social-science questions, and how closely do its automated judgments align with expert human assessments?}
On the social-science subset, our framework transfers robustly beyond STEM, and fine-grained prompting further strengthens both question quality and cognitive alignment. Under CoT, all six models already yield almost perfectly readable questions (100\% for all models) with high uniqueness and answerability (typically above 95\%), while FGP pushes these metrics close to saturation—for example, the answerability of Baichuan2-7B and Qwen2.5-7B rises from 95--96\% to 99--100\% (Table~\ref{tab:social_usability}). In terms of cognitive shifts, FGP systematically suppresses cognitive regression and increases cognitive leap on the same data, with CogReg dropping to 0\% for InternLM3-8B and Spark3.5-Max and CogLeap for Spark3.5-Max increasing from 42.50\% to 66.67\% (Table~\ref{tab:social_shift_distribution}), indicating that higher-order cognitive alignment can likewise be promoted via prompt design in non-STEM domains. Finally, the automatic rubric attains average agreement rates of 90.75\%, 98.08\%, and 95.83\% with expert annotations on uniqueness, readability, and answerability, respectively, while Bloom-level agreement is more modest at 46.58\% (Table~\ref{tab:social_human_auto_full}), suggesting that our framework is reliable for cross-domain automatic screening and quality control, yet fine-grained estimation of cognitive depth remains a challenging open problem.

\begin{table}[t]
\centering
\footnotesize
\caption{Performance of different LLMs on question uniqueness, readability, and answerability on the social-science subset under Chain-of-Thought (CoT) and Fine-Grained prompting (FGP).}
\label{tab:social_usability}
\renewcommand{\arraystretch}{1.1}
\setlength{\tabcolsep}{2.5pt}
\begin{tabular}{lccc|ccc}
\toprule
\multirow{2}{*}{Model} & \multicolumn{3}{c}{Social Sci. \& COT} & \multicolumn{3}{c}{Social Sci. \& FGP} \\
\cmidrule(lr){2-4} \cmidrule(lr){5-7}
& Unique & Readable & Answerable & Unique & Readable & Answerable \\
\midrule
GLM-4-9B       & 98.00 & 100.00 & 100.00 & 100.00 & 100.00 & 99.00 \\
Qwen2.5-7B     & 99.00 & 100.00 & 96.00  & 100.00 & 100.00 & 99.00 \\
Baichuan2-7B   & 100.00 & 100.00 & 95.00 & 100.00 & 100.00 & 100.00 \\
Meta LLaMA3-8B & 85.00 & 100.00 & 97.00  & 91.00  & 99.00  & 93.00 \\
InternLM3-8B   & 100.00 & 100.00 & 100.00 & 100.00 & 100.00 & 100.00 \\
Spark3.5-Max   & 100.00 & 100.00 & 99.00  & 100.00 & 100.00 & 98.00 \\
\bottomrule
\end{tabular}
\end{table}

\begin{table}[t]
\centering
\footnotesize
\caption{Cognitive shift distributions (regression, drift, leap) across LLMs on the social-science subset under Chain-of-Thought (CoT) and Fine-Grained prompting (FGP). Reported values are percentages among instances with mismatched Bloom levels.}
\label{tab:social_shift_distribution}
\renewcommand{\arraystretch}{1.1}
\setlength{\tabcolsep}{2.5pt}
\begin{tabular}{lccc|ccc}
\toprule
\multirow{2}{*}{Model} & \multicolumn{3}{c}{Social Sci. \& COT} & \multicolumn{3}{c}{Social Sci. \& FGP} \\
\cmidrule(lr){2-4} \cmidrule(lr){5-7}
& CogReg & CogDrift & CogLeap & CogReg & CogDrift & CogLeap \\
\midrule
GLM-4-9B       & 8.47  & 54.24 & 37.29 & 3.23  & 48.39 & 48.39 \\
Qwen2.5-7B     & 39.13 & 36.96 & 23.91 & 0.00  & 69.57 & 30.43 \\
Baichuan2-7B   & 18.42 & 44.74 & 36.84 & 16.18 & 35.29 & 48.53 \\
Meta LLaMA3-8B & 37.04 & 29.63 & 33.33 & 33.96 & 30.19 & 35.85 \\
InternLM3-8B   & 18.18 & 41.56 & 40.26 & 0.00  & 50.00 & 50.00 \\
Spark3.5-Max   & 2.50  & 55.00 & 42.50 & 0.00  & 33.33 & 66.67 \\
\bottomrule
\end{tabular}
\end{table}

\begin{table}[t]
\centering
\footnotesize
\caption{Agreement (\%) between automatic metrics and human annotations on the social-science subset under Chain-of-Thought (CoT) and Fine-Grained prompting (FGP).}
\label{tab:social_human_auto_full}
\renewcommand{\arraystretch}{1.05}
\setlength{\tabcolsep}{4pt}
\begin{tabular}{llcccc}
\toprule
Model & Prompt & Bloom & Unique & Readable & Answerable \\
\midrule
Baichuan2-7B   & CoT & 32.00 & 87.00 & 99.00 & 93.00  \\
GLM-4-9B       & CoT & 38.00 & 88.00 & 99.00 & 98.00 \\
InternLM3-8B   & CoT & 25.00 & 97.00 & 97.00 & 95.00 \\
Meta LLaMA3-8B & CoT & 37.00 & 67.00 & 93.00 & 89.00 \\
Qwen2.5-7B     & CoT & 57.00 & 84.00 & 100.00 & 96.00 \\
Spark3.5-Max   & CoT & 52.00 & 93.00 & 100.00 & 99.00 \\
Baichuan2-7B   & FGP & 27.00 & 98.00 & 100.00 & 100.00 \\
GLM-4-9B       & FGP & 68.00 & 99.00 & 100.00 & 99.00 \\
InternLM3-8B   & FGP & 47.00 & 97.00 & 100.00 & 98.00 \\
Meta LLaMA3-8B & FGP & 36.00 & 80.00 & 90.00 & 88.00 \\
Qwen2.5-7B     & FGP & 75.00 & 99.00 & 100.00 & 98.00  \\
Spark3.5-Max   & FGP & 65.00 & 100.00 & 99.00 & 97.00 \\
\midrule
Avg            & --  & 46.58 & 90.75 & 98.08 & 95.83 \\
\bottomrule
\end{tabular}
\end{table}

\section{Conclusion and Future Work}
We present a cognition-aware evaluation framework for educational question generation, assessing large pretrained language models on Bloom-level alignment, knowledge coverage, and cognitive shift intensity. Using two structured prompting strategies, we conduct large-scale experiments on 20{,}700 questions and highlight three key findings:
\begin{itemize}[leftmargin=*]
\item \textbf{Structured prompting improves cognitive control.} Fine-Grained Prompting (FGP) yields higher Bloom-level consistency and lower redundancy than Chain-of-Thought (CoT).
\item \textbf{Models overproduce higher-order questions.} Most models drift toward higher-order questions despite constraints, revealing an abstraction bias that may overload learners.
\item \textbf{Recognition and generation remain decoupled.} Strong Bloom-level classification does not guarantee generation at the intended cognitive level.
\end{itemize}

Our framework integrates cognitive alignment, knowledge tracking, and shift analysis into an automated evaluation suite that supports fine-grained model diagnosis and informs curriculum design, instruction tuning, and adaptive learning.

\textbf{Future Work.} We plan to conduct classroom-based studies to assess whether cognitive shift metrics correlate with learning outcomes. We also aim to explore semi-structured generation—specifying cognitive goals to promote linguistic diversity—and extend the framework to open-ended domains such as the humanities via rubric-based scoring and reasoning-quality modeling. Our code is available at: \url{https://github.com/wmm228/LLM-Eval}.

\bibliographystyle{ACM-Reference-Format}
\bibliography{sample-base}

@inproceedings{wang2023selfinstruct,
  title     = {Self-Instruct: Aligning Language Models with Self-Generated Instructions},
  author    = {Wang, Yizhong and Kordi, Yeganeh and Mishra, Swaroop and Liu, Alisa
               and Smith, Noah A. and Khashabi, Daniel and Hajishirzi, Hannaneh},
  booktitle = {Proceedings of the 61st Annual Meeting of the Association for
               Computational Linguistics (ACL)},
  year      = {2023}
}

@inproceedings{xu2025magpie,
  title     = {Magpie: Alignment Data Synthesis from Scratch by Prompting Aligned
               LLMs with Nothing},
  author    = {Xu, Zhangchen and Jiang, Fengqing and Niu, Luyao and Deng, Yuntian
               and Poovendran, Radha and Choi, Yejin and Lin, Bill Yuchen},
  booktitle = {Proceedings of the International Conference on Learning
               Representations (ICLR)},
  year      = {2025}
}

@inproceedings{constantinides2025autoq,
  title     = {Auto-Q: Automated Domain Questions Generation for Industrial Assets},
  author    = {Constantinides, Christodoulos and Sharma, Varun and Lin, Sheng
               and Zhou, Nan and Chaudhury, Bhaskar and Patel, Divya},
  booktitle = {Proceedings of the {AAAI} Conference on Artificial Intelligence},
  year      = {2025}
}

@inproceedings{scaria2024automated,
  title={Automated Educational Question Generation at Different Bloom’s Skill Levels Using Large Language Models: Strategies and Evaluation},
  author={Scaria, N and Chenna, Dharani S and Subramani, D},
  booktitle={International Conference on Artificial Intelligence in Education},
  pages={165--179},
  year={2024},
  organization={Springer Nature Switzerland},
  address={Cham}
}

@inproceedings{chen2024empowering,
  title={Empowering private tutoring by chaining large language models},
  author={Chen, Y and Ding, N and Zheng, HT and others},
  booktitle={Proceedings of the 33rd ACM International Conference on Information and Knowledge Management},
  pages={354--364},
  year={2024}
}

@inproceedings{kokku2018augmenting,
  title={Augmenting classrooms with AI for personalized education},
  author={Kokku, R and Sundararajan, S and Dey, P and others},
  booktitle={2018 IEEE International Conference on Acoustics, Speech and Signal Processing (ICASSP)},
  pages={6976--6980},
  year={2018},
  organization={IEEE}
}

@inproceedings{mucciaccia2025automatic,
  title={Automatic Multiple-Choice Question Generation and Evaluation Systems Based on LLM: A Study Case With University Resolutions},
  author={Mucciaccia, SS and Paixão, TM and Mutz, FW and others},
  booktitle={Proceedings of the 31st International Conference on Computational Linguistics},
  pages={2246--2260},
  year={2025}
}

@inproceedings{gong2022khanq,
  title={Khanq: A dataset for generating deep questions in education},
  author={Gong, H and Pan, L and Hu, H},
  booktitle={Proceedings of the 29th International Conference on Computational Linguistics},
  pages={5925--5938},
  year={2022}
}

@inproceedings{elkins2024how,
  title={How Teachers Can Use Large Language Models and Bloom’s Taxonomy to Create Educational Quizzes},
  author={Elkins, S and Kochmar, E and Cheung, JCK and others},
  booktitle={Proceedings of the AAAI Conference on Artificial Intelligence},
  volume={38},
  number={21},
  pages={23084--23091},
  year={2024}
}

@article{laban2022quiz,
  title={Quiz design task: Helping teachers create quizzes with automated question generation},
  author={Laban, P and Wu, CS and Murakhovs’ ka, L and others},
  journal={arXiv preprint arXiv:2205.01730},
  year={2022}
}

@inproceedings{muse2023pre,
  title={Pre-training with scientific text improves educational question generation (student abstract)},
  author={Muse, H and Bulathwela, S and Yilmaz, E},
  booktitle={Proceedings of the AAAI Conference on Artificial Intelligence},
  volume={37},
  number={13},
  pages={16288--16289},
  year={2023}
}

@article{yuan2022selecting,
  title={Selecting better samples from pre-trained LLMs: A case study on question generation},
  author={Yuan, X and Wang, T and Wang, YH and others},
  journal={arXiv preprint arXiv:2209.11000},
  year={2022}
}

@inproceedings{li2024planning,
  title={Planning First, Question Second: An LLM-Guided Method for Controllable Question Generation},
  author={Li, K and Zhang, Y},
  booktitle={Findings of the Association for Computational Linguistics ACL 2024},
  pages={4715--4729},
  year={2024}
}

@article{chen2024dr,
  title={Dr. academy: A benchmark for evaluating questioning capability in education for large language models},
  author={Chen, Y and Wu, C and Yan, S and others},
  journal={arXiv preprint arXiv:2408.10947},
  year={2024}
}

@article{kojima2022large,
  title={Large language models are zero-shot reasoners},
  author={Kojima, T and Gu, SS and Reid, M and others},
  journal={Advances in neural information processing systems},
  volume={35},
  pages={22199--22213},
  year={2022}
}

@article{shi2023prompt,
  title={Prompt space optimizing few-shot reasoning success with large language models},
  author={Shi, F and Qing, P and Yang, D and others},
  journal={arXiv preprint arXiv:2306.03799},
  year={2023}
}

@article{wei2022chain,
  title={Chain-of-thought prompting elicits reasoning in large language models},
  author={Wei, J and Wang, X and Schuurmans, D and others},
  journal={Advances in neural information processing systems},
  volume={35},
  pages={24824--24837},
  year={2022}
}

@book{anderson2001taxonomy,
  title={A taxonomy for learning, teaching, and assessing: A revision of Bloom's taxonomy of educational objectives: complete edition},
  author={Anderson, LW and Krathwohl, DR},
  year={2001},
  publisher={Addison Wesley Longman, Inc.}
}

@inproceedings{tobler2022impact,
  title={The impact of prior knowledge in narrative-based learning on understanding biological concepts in higher education},
  author={Tobler, S and Sinha, T and Koehler, K and others},
  booktitle={Proceedings of the Annual Meeting of the Cognitive Science Society},
  volume={44},
  number={44},
  year={2022}
}

@inproceedings{bartel2023applying,
  title={Applying cognitive learning principles to practice: Challenges in translation and large-scale study design},
  author={Bartel, A and Matlen, B and Rohrer, D and others},
  booktitle={Proceedings of the Annual Meeting of the Cognitive Science Society},
  volume={45},
  number={45},
  year={2023}
}

@inproceedings{dsilva2023embedding,
  title={Embedding Equitable Research Practices into the Rigorous Study of a Cognitive Learning Intervention},
  author={D'Silva, K and Matlen, B},
  booktitle={Proceedings of the Annual Meeting of the Cognitive Science Society},
  volume={45},
  number={45},
  year={2023}
}

@inproceedings{arvan2023linguistic,
  title={Linguistic Cognitive Load Analysis on Dialogues with an Intelligent Virtual Assistant},
  author={Arvan, M and Valizadeh, M and Haghighat, P and others},
  booktitle={Proceedings of the Annual Meeting of the Cognitive Science Society},
  volume={45},
  number={45},
  year={2023}
}

@article{daheim2024stepwise,
  title={Stepwise verification and remediation of student reasoning errors with large language model tutors},
  author={Daheim, N and Macina, J and Kapur, M and others},
  journal={arXiv preprint arXiv:2407.09136},
  year={2024}
}

@article{wang2024book2dial,
  title={Book2Dial: Generating Teacher-Student Interactions from Textbooks for Cost-Effective Development of Educational Chatbots},
  author={Wang, J and Macina, J and Daheim, N and others},
  journal={arXiv preprint arXiv:2403.03307},
  year={2024}
}

@inproceedings{chen2022grow,
  title={Grow-and-Clip: Informative-yet-Concise Evidence Distillation for Answer Explanation},
  author={Chen, Y and Xiao, Y and Liu, B},
  booktitle={2022 IEEE 38th International Conference on Data Engineering (ICDE)},
  pages={741--754},
  year={2022},
  organization={IEEE}
}

@inproceedings{fedyk2023how,
  title={How to Leverage Machine Learning Interpretability and Explainability to Generate Hypotheses in Cognitive Psychology},
  author={Fedyk, M and Ray, M},
  booktitle={Proceedings of the Annual Meeting of the Cognitive Science Society},
  volume={45},
  number={45},
  year={2023}
}

@inproceedings{ravikiran2025teemil,
  title={TEEMIL: Towards Educational MCQ Difficulty Estimation in Indic Languages},
  author={Ravikiran, M and Vohra, S and Verma, R and others},
  booktitle={Proceedings of the 31st International Conference on Computational Linguistics},
  pages={2085--2099},
  year={2025}
}

@inproceedings{qiu2025knowledge,
  title={A Knowledge Graph Reasoning-Based Model for Computerized Adaptive Testing},
  author={Qiu, X and Chen, Z},
  booktitle={Proceedings of the 31st International Conference on Computational Linguistics},
  pages={5295--5304},
  year={2025}
}

@inproceedings{liu2025from,
  title={From Superficial to Deep: Integrating External Knowledge for Follow-up Question Generation Using Knowledge Graph and LLM},
  author={Liu, J and Huang, Y and Bi, S and others},
  booktitle={Proceedings of the 31st International Conference on Computational Linguistics},
  pages={828--840},
  year={2025}
}

@inproceedings{feng2025rgr,
  title={RGR-KBQA: Generating Logical Forms for Question Answering Using Knowledge-Graph-Enhanced Large Language Model},
  author={Feng, T and He, L},
  booktitle={Proceedings of the 31st International Conference on Computational Linguistics},
  pages={3057--3070},
  year={2025}
}

@inproceedings{liao2025skintern,
  title={SKIntern: Internalizing Symbolic Knowledge for Distilling Better CoT Capabilities into Small Language Models},
  author={Liao, H and He, S and Hao, Y and others},
  booktitle={Proceedings of the 31st International Conference on Computational Linguistics},
  pages={3203--3221},
  year={2025}
}

@inproceedings{liao2025awakening,
  title={Awakening Augmented Generation: Learning to Awaken Internal Knowledge of Large Language Models for Question Answering},
  author={Liao, H and He, S and Xu, Y and others},
  booktitle={Proceedings of the 31st International Conference on Computational Linguistics},
  pages={1333--1352},
  year={2025}
}

@inproceedings{park2024large,
  title={Large Language Models are Students at Various Levels: Zero-shot Question Difficulty Estimation},
  author={Park, JW and Park, SJ and Won, HS and others},
  booktitle={Findings of the Association for Computational Linguistics: EMNLP 2024},
  pages={8157--8177},
  year={2024}
}

@article{zaman2024dataset,
  title={Dataset of computer science course queries from students: Categorized and scored according to Bloom's taxonomy},
  author={Zaman, Khandoker Ashik Uz and Islam, Ashraful and Islam, Yusuf Mahbubul and Sayed, Md Abu},
  journal={Data in Brief},
  volume={53},
  pages={110109},
  year={2024},
  publisher={Elsevier}
}

@misc{du2024glm4,
  title        = {ChatGLM: A Family of Large Language Models from GLM‑130B to GLM‑4 All Tools},
  author       = {Zhengxiao Du and et al.},
  year         = {2024},
  eprint       = {2406.12793},
  archivePrefix= {arXiv},
  primaryClass = {cs.CL}
}

@misc{yang2025qwen25,
  title        = {Qwen2.5 Technical Report},
  author       = {An Yang and et al.},
  year         = {2025},
  eprint       = {2412.15115},
  archivePrefix= {arXiv},
  primaryClass = {cs.CL}
}

@misc{baichuan2023,
  title        = {Baichuan2: Open Large Language Models},
  author       = {Baichuan Inc.},
  year         = {2023},
  eprint       = {2310.11453},
  archivePrefix= {arXiv},
  primaryClass = {cs.CL}
}

@misc{internlm2024,
  title        = {InternLM2 Technical Report},
  author       = {Zheng Cai and et al.},
  year         = {2024},
  eprint       = {2403.17297},
  archivePrefix= {arXiv},
  primaryClass = {cs.CL}
}

@misc{dubey2024llama3,
  title        = {The Llama 3 Herd of Models},
  author       = {Dubey et al. and LLaMA Team},
  year         = {2024},
  eprint       = {2407.21783},
  archivePrefix= {arXiv},
  primaryClass = {cs.AI}
}

\appendix

\section{Comparative dimensions.}
In Table~\ref{tab:rw_comparison}, DS denotes explicit data synthesis for instructions or questions, Bloom explicit control/annotation of Bloom levels, Cog explicit evaluation of cognitive depth or level shifts, Cov measurement of coverage of target knowledge units, Human systematic human judgments, and Edu a primary focus on educational question generation or tutoring. A ``\checkmark'' indicates that the aspect is explicitly addressed, while ``--'' means it is not a primary focus.

\section{Usable question counts.}
After all validity and usability filters, we obtain the following per-model question counts. \emph{Computer science} (CoT / FGP): GLM 1278 / 1406, Qwen 1054 / 1399, Baichuan 1003 / 1174, LLaMA 1331 / 1372, InternLM 1332 / 1396, Spark 1370 / 1336. \emph{K--12 mathematics} (FGP only): Baichuan 419, GLM 427, InternLM 397, LLaMA 413, Qwen 422, Spark 426. \emph{Social science} human-evaluation subset (CoT / FGP): Baichuan 95 / 100, GLM 98 / 99, InternLM 100 / 100, LLaMA 82 / 85, Qwen 95 / 99, Spark 99 / 98. The remaining preprocessing and filtering details will be documented in our public GitHub repository.

\onecolumn 
\section{Classification Performance}

\begin{figure}[htbp]    
\centering    
\includegraphics[width=0.9\linewidth]{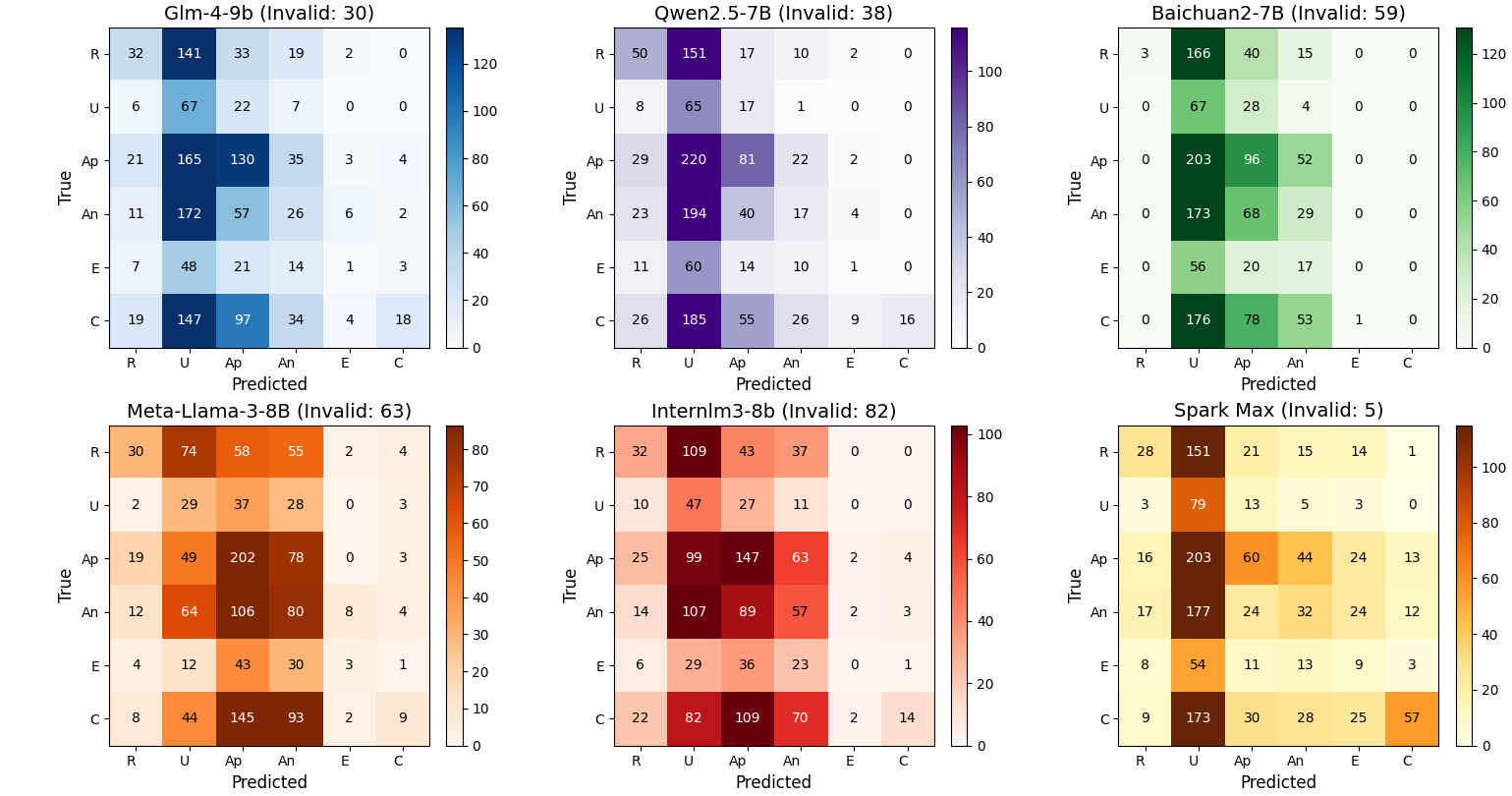}    
\caption{Classification performance of different models across Bloom’s taxonomy categories. The "Application" (Ap) category exhibits higher classification accuracy, whereas misclassification is more prevalent in "Analysis" (An) and "Creation" (C).}    
\label{fig:class}
\end{figure}

\section{Knowledge Unit}

\begin{table}[htbp]
\centering
\small
\caption{Mapping between datasets and knowledge topics.}
\label{tab:dataset_topics}
\begin{tabular}{l p{10cm}}
\toprule
\textbf{Dataset} & \textbf{Knowledge Topics} \\
\midrule
Computer Science & 
Binary Search Tree; 
Doubly Linked List; 
Graph; 
Hashing; 
Lists; 
Nodes; 
Pointers; 
Python Review; 
Queues; 
Recursion; 
Singly Linked List; 
Stacks \\
\midrule
K-12 Math & 
Total cost problems -- one-step; 
Find volume given base perimeter and surface area; 
Find length, width, height from volume; 
Base area of a rectangular prism; 
Cube coloring problems; 
Applications of prime numbers; 
Inverse proportion problems -- tiling with square bricks; 
Find height of cylinder and cone; 
Solve fraction problems by tracking invariants; 
Volume calculation of a cylinder; 
Positive and negative numbers -- expressing relative quantities; 
Identifying properties of lines, rays, and segments; 
Quadrants of a directly proportional function -- find variable range; 
Concepts related to cones; 
Apply periodicity to compute function values; 
Angles with terminal sides on the same line; 
Use derivative of a tangent curve to find parameters; 
Number of intersection points between a line and a hyperbola; 
Exponent strategies via permutation problems; 
Application of combination formulas -- evaluation \\
\midrule
Social Science &
Group Processes \& Organizational Behavior; 
Attitudes, Persuasion \& Social Influence; 
Inequality, Class \& Mobility; 
Culture, Norms \& Socialization; 
Causal Inference \& Research Design \\
\bottomrule
\end{tabular}
\end{table}

\section{Prompt Template}

\begin{figure}[htbp]
    \centering
    \includegraphics[width=0.72\linewidth]{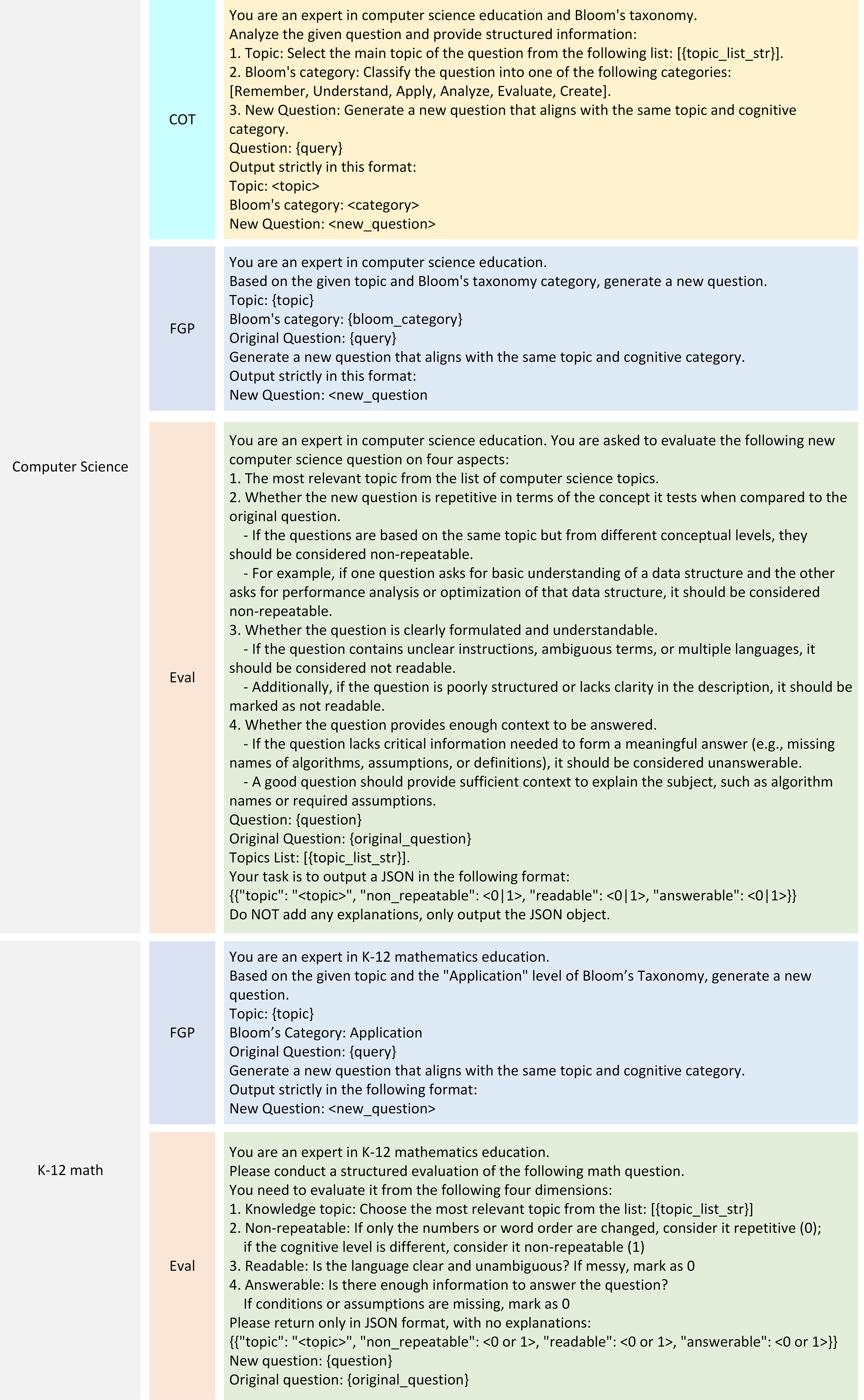}  
    \caption{Prompt templates for guiding LLMs to generate and evaluate educational questions in computer science and K-12 mathematics.}
    \label{fig:prompt}  
\end{figure}


\end{document}